\documentstyle[12pt]{article}
\def\ie{{\em i.e.}}
\def\eg{{\em e.g.}}

\def\beq{\begin{equation}}
\def\eeq{\end{equation}}

\catcode`\@=11 
\def\coeff#1#2{{\textstyle{#1\over #2}}}

\def\lsim{\mathrel{\mathpalette\@versim<}}
\def\gsim{\mathrel{\mathpalette\@versim>}}
\def\@versim#1#2{\vcenter{\offinterlineskip
    \ialign{$\m@th#1\hfil##\hfil$\crcr#2\crcr\sim\crcr } }}
\def\etal{{\em et. al.}}
\def\JL{J. L. Lopez}
\def\DVN{D. V. Nanopoulos}

\def\r#1{$\bf#1$}
\def\rb#1{$\bf\overline{#1}$}

\def\t1{{\tilde 1}}

\def\GeV{\,{\rm GeV}}

\def\to{\rightarrow}

\def\NPB#1#2#3{Nucl. Phys. B {\bf#1} (19#2) #3}
\def\PLB#1#2#3{Phys. Lett. B {\bf#1} (19#2) #3}

\def\PRD#1#2#3{Phys. Rev. D {\bf#1} (19#2) #3}
\def\PRL#1#2#3{Phys. Rev. Lett. {\bf#1} (19#2) #3}
\def\PRT#1#2#3{Phys. Rep. {\bf#1} (19#2) #3}

\def\TAMU#1{Texas A \& M University preprint CTP-TAMU-#1}

\begin{document}
\begin{flushright}
\baselineskip=12pt
{CERN-TH.6986/93}\\
{CTP-TAMU-44/93}\\
{ACT-17/93}\\
\end{flushright}

\begin{center}
\vglue 1cm
{\Large\bf Large $\bf(g-2)_\mu$ in SU(5)xU(1) Supergravity Models\\}
\vglue 1cm
{JORGE L. LOPEZ$^{(a),(b)}$, D. V. NANOPOULOS$^{(a),(b),(c)}$, and XU
WANG$^{(a),(b)}$\\}
\vglue 0.4cm
{\em $^{(a)}$Center for Theoretical Physics, Department of Physics, Texas A\&M
University\\}
{\em College Station, TX 77843--4242, USA\\}
{\em $^{(b)}$Astroparticle Physics Group, Houston Advanced Research Center
(HARC)\\}
{\em The Woodlands, TX 77381, USA\\}
{\em $^{(c)}$CERN, Theory Division, 1211 Geneva 23, Switzerland\\}
\baselineskip=12pt

\vglue 1cm
{\tenrm ABSTRACT}
\end{center}
{\rightskip=3pc
 \leftskip=3pc
\noindent
We compute the supersymmetric contribution to the anomalous magnetic moment of
the muon within the context of $SU(5)\times U(1)$ supergravity models. The
largest possible contributions to $a^{susy}_\mu$ occur for the largest allowed
values of $\tan\beta$ and can easily exceed the present experimentally allowed
range, even after the LEP lower bounds on the sparticle masses are imposed.
Such $\tan\beta$ enhancement implies that $a^{susy}_\mu$ can greatly exceed
both the electroweak contribution ($\approx1.95\times10^{-9}$) and the present
hadronic uncertainty ($\approx\pm1.75\times10^{-9}$). Therefore, the new E821
Brookhaven experiment (with an expected accuracy of $0.4\times10^{-9}$) should
explore a large fraction (if not all) of the parameter space of these models,
corresponding to slepton, chargino, and squarks masses as high as 200, 300, and
1000 GeV respectively. Moreover, contrary to popular belief, the $a^{susy}_\mu$
contribution can have either sign, depending on the sign of the Higgs mixing
parameter $\mu$: $a^{susy}_\mu>0\,(<0)$ for $\mu>0$ ($\mu<0$).
The present $a_\mu$ constraint excludes chargino masses in the range
$45-120\GeV$ depending on the value of $\tan\beta$, although there are no
constraints for $\tan\beta\lsim8$. We also compute $a^{susy}_\tau$ and find
$|a^{susy}_\tau|\approx(m_\tau/m_\mu)^2\,|a^{susy}_\mu|\lsim10^{-5}$ and
briefly comment on its possible observability.
}
\vspace{0.5in}
\begin{flushleft}
\baselineskip=12pt
{CERN-TH.6986/93}\\
{CTP-TAMU-44/93}\\
{ACT-17/93}\\
August 1993
\end{flushleft}
\vfill\eject
\setcounter{page}{1}
\pagestyle{plain}

\baselineskip=14pt

\section{Introduction}
The experimental measurements of the leptonic anomalous magnetic moments have
been carried out to such great accuracy that their agreement with the
theoretical calculations has been one of the most spectacular successes of
quantum field theory and QED in particular \cite{kinoI}. While efforts to go to
higher experimental accuracies are being pursued mainly to test the standard
model electroweak contribution to $a_\mu\equiv{1\over2}(g-2)_\mu$ \cite{newg},
new physics might come into play as well. Since realistic supersymmetric models
predict a sparticle mass spectrum which can be as light as $\approx45\GeV$, it
is possible that an experimental measurement of the muonic $g-2$ factor with
high accuracy could put some constraints on the new and yet-to-be-found
sparticles, and therefore on the parameter space on the various supersymmetric
models.

The long standing experimental values of $a_\mu$ for each sign of the muon
electric charge \cite{oldg} can be averaged to yield \cite{kinoII}
\beq
a^{exp}_\mu=1\ 165\ 923(8.5)\times10^{-9}.
\eeq
The uncertainty on the last digit is indicated in parenthesis. On the other
hand, the various standard model contributions to $a_\mu$ have been estimated
to be as follows \cite{kinoII}
\begin{eqnarray}
{\rm QED:}&&1\ 165\ 846\ 984(17)(28)\times10^{-12}\\
{\rm had.1:}&&7\ 068(59)(164)\times10^{-11}\\
{\rm had.2:}&&-90(5)\times10^{-11}\\
{\rm had.3:}&&49(5)\times10^{-11}\\
{\rm Total\ hadronic:}&&7\ 027(175)\times10^{-11}\\
{\rm Electroweak:}&&195(10)\times10^{-11}
\end{eqnarray}
The total standard model prediction is then \cite{kinoII}
\beq
a^{SM}_\mu=116\ 591\ 9.20(1.76)\times10^{-9}.
\eeq
Subtracting the experimental result gives \cite{kinoII}
\beq
a^{SM}_\mu-a^{exp}_\mu=-3.8(8.7)\times10^{-9},\label{diff}
\eeq
which is perfectly consistent with zero. The uncertainty in the theoretical
prediction is dominated by the uncertainty in the lowest order hadronic
contribution (had.1), which ongoing experiments at Novosibirsk hope to reduce
by a factor of two in the near future. This is an important preliminary step
to testing the electroweak contribution, which is of the same order. The
uncertainty in the experimental determination of $a_\mu$ is expected to be
reduced significantly (down to $0.4\times10^{-9}$) by the new E821 Brookhaven
experiment \cite{newg}, which is scheduled to start taking data in late 1994.
Any beyond-the-standard-model contribution to $a_\mu$ (with presumably
negligible uncertainty) will simply be added to the central value in
Eq.~(\ref{diff}). Therefore, we can obtain an allowed interval for any
supersymmetric contribution, such that $a^{susy}_\mu+a^{SM}_\mu-a^{exp}_\mu$ is
consistent with zero at some given confidence level,
\begin{eqnarray}
-4.9\times10^{-9}&<a^{susy}_\mu&<12.5\times10^{-9},\qquad{\rm at\ 1\sigma};\\
-10.5\times10^{-9}&<a^{susy}_\mu&<18.1\times10^{-9},\qquad{\rm at\ 90\%CL};\\
-13.2\times10^{-9}&<a^{susy}_\mu&<20.8\times10^{-9},\qquad{\rm at\
95\%CL}.\label{bounds}
\end{eqnarray}

The supersymmetric contributions to $a_\mu$ have been computed to various
degrees of completeness and in the context of several models, including
the minimal supersymmetric standard model (MSSM)
\cite{Fayet,GM,KKS,YACN,Vendramin}, an $E_6$ string-inspired model \cite{E6g},
and a non-minimal MSSM with an additional singlet \cite{NMSSMg,abel}. Because
of the large number of parameters appearing in the typical formula for
$a^{susy}_\mu$, various contributions have often been neglected and numerical
results are basically out of date. More importantly, a contribution which is
roughly proportional to the ratio of Higgs vacuum expectation values
($\tan\beta$), even though known for a while \cite{KKS,YACN,Vendramin,abel},
has to date remained greatly unappreciated. This has been the case because in
the past only small values of $\tan\beta$ were usually considered and the
enhancement of $a^{susy}_\mu$, which is the focus of this paper, was not
evident. In fact, such enhancement can easily make $a^{susy}_\mu$ run in
conflict with the bounds given in Eq.~(\ref{bounds}), even after the LEP lower
bounds on the sparticle masses are imposed.

In this paper we compute $a^{susy}_\mu$ in the context of supergravity models
based on the $SU(5)\times U(1)$ (flipped $SU(5)$) gauge group \cite{EriceDec92}
supplemented by two string-inspired (the so-called no-scale \cite{LNZI} and
dilaton \cite{LNZII}) soft-supersymmetry-breaking scenarios. Among the various
interesting properties these models possess, perhaps the most relevant ones
here are that their sparticle mass spectra are as light as could possibly be
for a supergravity model, and that large values of $\tan\beta$ are typical.
Moreover, the complete parameter space can be described in terms of only three
variables: the top-quark mass ($m_t$), $\tan\beta$, and the gluino mass
($m_{\tilde g}$). Thus, we are able to produce quite specific predictions for
$a^{susy}_\mu$ throughout the parameter space. This paper is in line with a
series of phenomenological calculations which the authors have performed
recently within the context of this class of models \cite{phenom}.

Besides the experiments on $(g-2)_\mu$, there have been suggestions
\cite{taug,SLM} that the anomalous magnetic moment of the tau lepton could be
measured at hadron supercolliders to the precision of $\approx 10^{-5}$.
Therefore, we also compute $a^{susy}_\tau$ in these models.

\section{The flipped SU(5) supergravity models}
The models of interest in this paper are based on the gauge group $SU(5)\times
U(1)$ and have the property of gauge coupling unification at the scale
$10^{18}\GeV$ \cite{EriceDec92}. This implies that their matter content must
include additional particles beyond the supersymmetric standard model with two
Higgs doublets, otherwise the unification scale would be $10^{16}\GeV$. Indeed,
an extra pair of vector-like quark doublets with mass $\sim10^{12}\GeV$ and
a pair of charge $-1/3$ quark singlets with mass $\sim10^6\GeV$ appear in the
spectrum. This additional particles form complete \r{10},\rb{10} $SU(5)\times
U(1)$ multiplets and are seen to occur in a string-derived version of this
model \cite{LNY}. The unification scale is also consistent with that expected
in string models of this kind \cite{Lacaze}. Besides contributing to the gauge
coupling beta functions for scales above their masses, the new particles do not
have any other noticeable effects. Nonetheless, such subtle changes in slope
propagate throughout the whole system of renormalization group equations for
the gauge and Yukawa couplings, as well as the scalar masses and trilinear
scalar couplings. An effect of similar magnitude is a consequence of the
``extra" running down from $10^{18}\GeV$  relative to a model which unifies at
$10^{16}\GeV$.

This class of supergravity models can be described completely in terms of just
three parameters: (i) the top-quark mass ($m_t$), (ii) the ratio of Higgs
vacuum expectation values, which satisfies $1\lsim\tan\beta\lsim40$, and (iii)
the gluino mass, which is cut off at 1 TeV. This simplification in the number
of input parameters is possible because of specific scenarios for the universal
soft-supersymmetry-breaking parameters ($m_0,m_{1/2},A$) at the unification
scale. These three parameters can be computed in specific string models in
terms of just one of them \cite{IL}. In the no-scale model one obtains
$m_0=A=0$, whereas in the dilaton model the result is
$m_0=\frac{1}{\sqrt{3}}m_{1/2}, A=-m_{1/2}$.
After running the renormalization group equations from high to low energies,
at the low-energy scale the requirement of radiative electroweak symmetry
breaking introduces two further constraints which among other things determine
the magnitude of the Higgs mixing term $\mu$, although its sign remains
undetermined. Finally, all the known phenomenological constraints on the
sparticle masses are imposed (most importantly the chargino, slepton, and Higgs
mass bounds). This procedure is well documented in the literature
\cite{aspects} and yields the allowed parameter spaces for the no-scale
\cite{LNZI} and dilaton \cite{LNZII} cases.

These allowed parameter spaces in the three defining variables
($m_t,\tan\beta,m_{\tilde g}$) consist of a discrete set of points for
three values of $m_t$ ($m_t=130,150,170\GeV$), and a discrete set of allowed
values for $\tan\beta$, starting at 2\footnote{Note that $\tan\beta>1$ is
required by the radiative breaking mechanism, and the LEP lower bound on the
lightest Higgs boson mass ($m_h\gsim60\GeV$ \cite{LNPWZh}) is quite
constraining for $1<\tan\beta<2$.} and running (in steps of two) up to 32 (46)
for the no-scale (dilaton) case. The allowed values of $m_{\tilde g}$ vary
from a minimum value of $\approx200\GeV$ up to 1 TeV, depending on the value of
$\tan\beta$. For each of these points in parameter space there corresponds one
set of sparticle and Higgs masses, as well as various diagonalizing matrices
for the neutralino, chargino, slepton, and squark masses. In particular, {\em
all} of the parameters that appear in the formula for $a^{susy}_\mu$ given
below can be obtained for any given point in parameter space.

\begin{table}
\hrule
\caption{The approximate proportionality coefficients to the gluino mass, for
the various sparticle masses in the two supersymmetry breaking scenarios
considered.}
\label{Table1}
\begin{center}
\begin{tabular}{|c|c|c|}\hline
&no-scale&dilaton\\ \hline
$\tilde e_R,\tilde \mu_R$&$0.18$&$0.33$\\
$\tilde\nu$&$0.18-0.30$&$0.33-0.41$\\
$2\chi^0_1,\chi^0_2,\chi^\pm_1$&$0.28$&$0.28$\\
$\tilde e_L,\tilde \mu_L$&$0.30$&$0.41$\\
$\tilde q$&$0.97$&$1.01$\\
$\tilde g$&$1.00$&$1.00$\\ \hline
\end{tabular}
\end{center}
\hrule
\end{table}

In the models we consider all sparticle masses scale with the gluino mass, with
a mild $\tan\beta$ dependence. In Table~\ref{Table1} we give the approximate
proportionality coefficient (to the gluino mass) for each sparticle mass. Note
that the relation $2m_{\chi^0_1}\approx m_{\chi^0_2}\approx m_{\chi^\pm_1}$
holds to good approximation. The third-generation squark and slepton masses
also scale with $m_{\tilde g}$, but the relationships are smeared by a strong
$\tan\beta$ dependence. From Table~\ref{Table1} one can (approximately)
translate any bounds on a given sparticle mass on bounds on all the other
sparticle masses.

\section{Calculation and discussion of results}
There are two sources of one-loop supersymmetric contributions to $a_\mu$: (i)
with neutralinos and smuons in the loop; and (ii) with charginos and sneutrinos
in the loop. In the former case it is necessary to diagonalize the smuon mass
matrix to get the mass eigenstates,
\beq
	DMD^\dagger={\rm diag}(m_{{\tilde \mu}_{1}}^2,m_{{\tilde \mu}_{2}}^2),
\eeq
where $M$ is the smuon mass matrix
\beq
	M=\left(\matrix{m_{{\tilde\mu} LL}^{2}&m_{{\tilde\mu}LR}^{2}\cr
			m_{{\tilde\mu}LR}^{2}&m_{{\tilde\mu}RR}^{2}\cr}\right),
\eeq
and $D$ is the orthogonal rotation matrix. This gives the mass eigenstates
\beq
	\tilde \mu_{i}=D_{i1}\tilde\mu_{L}+D_{i2}\tilde\mu_{R}, \qquad i=1,2.
\eeq
Therefore, the rotation angle can be expressed as
\beq
	\tan(2\theta)={{2m_{\tilde{\mu}LR}^{2}}\over{(m_{\tilde{\mu}_{LL}}^{2}-
m_{\tilde{\mu}_{RR}}^{2})}},
\eeq
where
\beq
	m_{\tilde\mu_{LR}}^{2}=m_{\mu}(A_{\mu}+\mu \tan\beta).
\eeq
It is clear that because of the smallness of the muon mass compared
with the sparticle mass scale, the mixing angle is quite small. The general
formula for the lowest order supersymmetric contribution to $a_\mu$ has been
given in Refs.~\cite{KKS,YACN,Vendramin,abel}. Here we use the expression in
Ref.~\cite{abel},
\begin{eqnarray}
a^{susy}_\mu=-\frac{g^2_2}{8\pi^2}
\Biggl\{\sum_{\chi^0_i,\tilde \mu_j} \frac{m_\mu}{m_{\chi^0_i}}
	&\biggl[&\!\!\!\!(-1)^{j+1}\sin(2\theta)B_1(\eta_{ij})\tan\theta_W N_{i1}
		[\tan\theta_W N_{i1}+N_{i2}]\nonumber\\
&+&\frac{m_\mu}{2M_W\cos\beta}
	B_1(\eta_{ij})N_{i3}[3\tan\theta_W N_{i1}+N_{i2}]\nonumber\\
&+&\left(\frac{m_{\mu}}{m_{\chi^0_i}}\right)^{2}A_1(\eta_{ij})
\Bigl\{\coeff{1}{4}[\tan\theta_W N_{i1}+N_{i2}]^{2}+
[\tan\theta_W N_{i1}]^{2}\Bigr\}\biggr]\nonumber\\
 &&\hskip-1.3in -\sum_{\chi_j^\pm}\left[\frac{m_{\mu}m_{\chi_j^\pm}}
{m_{\tilde \nu}^2}
\frac{m_{\mu}}{\sqrt{2}M_W\cos\beta} B_2(\kappa_j)V_{j1}U_{j2}+
\left(\frac{m_{\mu}}
{m_{\tilde\nu}}\right)^{2}\frac{A_1(\kappa_j)}{2}V^2_{j1}\right]\Biggr\}.
\label{formula}
\end{eqnarray}
where $N_{ij}$ are elements of the matrix which diagonalizes the neutralino
mass matrix, and $U_{ij},V_{ij}$ are the corresponding ones for the chargino
mass matrix, in the notation of Ref.~\cite{HK}. Also,
\beq
\eta_{ij}=\left[1
-\left(\frac{m_{\tilde\mu_j}}{m_{\chi^0_i}}\right)^2\right]^{-1},\qquad
\kappa_j=\left[1
-\left(\frac{m_{\chi^\pm_j}}{m_{\tilde\nu}}\right)^2\right]^{-1},
\eeq
and
\begin{eqnarray}
B_1(x)&=&x^2-\coeff{1}{2}x+x^2(x-1)\ln\left(\frac{x-1}{x}\right),\\
A_1(x)&=&x^3-\coeff{1}{2}x^2-\coeff{1}{6}x+x^3(x-1)\ln\left(\frac{x-1}{x}\right),\\
B_2(x)&=&-x^2-\coeff{1}{2}x-x^3\ln\left(\frac{x-1}{x}\right).
\end{eqnarray}

As has been pointed out, the mixing angle of the smuon eigenstates is small
(although it can be enhanced for large $\tan\beta$) and it makes the
neutralino-smuon contribution suppressed. Moreover, the various
neutralino-smuon contributions (the first three lines
in Eq.~(\ref{formula})) tend to largely cancel among themselves
\cite{Vendramin}.\footnote{The original Fayet formula \cite{Fayet} is obtained
from the third neutralino-smuon contribution in the limit of a massless
photino and no smuon mixing.} This means that the chargino-sneutrino
contributions (on the fourth line in Eq.~(\ref{formula})) will likely be the
dominant ones. In fact, as we stress in this paper, the first
chargino-sneutrino contribution (the ``gauge-Yukawa" contribution) is enhanced
relative to the second one (the ``pure gauge" contribution) for large values
of $\tan\beta$. This can be easily seen as follows.

Picturing the chargino-sneutrino one-loop diagram, with the photon being
emitted off the chargino line, there are two ways in which the helicity of the
muon can be flipped, as is necessary to obtain a non-vanishing $a_\mu$:
\begin{description}
\item (i) It can be flipped by an explicit muon mass insertion on one of the
external muon lines, in which case the coupling at the vertices is between a
left-handed muon, a sneutrino, and the wino component of the chargino and has
magnitude $g_2$. It then follows that $a_\mu$ will be proportional to
$g^2_2(m_\mu/\tilde m)^2 |V_{j1}|^2$, where $\tilde m$ is a supersymmetric mass
in the loop and the $V_{j1}$ factor picks out the wino component of the $j$-th
chargino. This is the origin of the ``pure gauge" contribution to
$a^{susy}_\mu$.
\item (ii) Another possibility is to use the muon Yukawa coupling on one of
the vertices, which flips the helicity and couples to the Higgsino component
of the chargino. One also introduces a chargino mass insertion to switch to the
wino component and couple with strength $g_2$ at the other vertex. The
contribution is now proportional to $g_2\lambda_\mu(m_\mu m_{\chi^\pm_j}/\tilde
m^2)V_{j1}U_{j2}$, where $U_{j2}$ picks out the Higgsino component of the
$j$-th chargino. The muon Yukawa coupling is given by $\lambda_\mu=g_2
m_\mu/(\sqrt{2}M_W\cos\beta)$. This is the origin of the gauge-Yukawa
contribution to $a^{susy}_\mu$.
\end{description}
The ratio of the ``pure gauge" to the ``gauge-Yukawa" contributions is
roughly then
\beq
g^2_2\,(m_\mu/\tilde m)/(g_2\lambda_\mu)\sim g_2/\sqrt{1+\tan^2\beta},
\eeq
for $\tilde m\sim100\GeV$. Thus, for small $\tan\beta$ both contributions
are comparable, but for large $\tan\beta$ the ``gauge-Yukawa" contribution
is greatly enhanced.\footnote{A similar enhancement in the second
neutralino-smuon contribution is suppressed by small Higgsino admixtures
(\ie, $|N_{13}|,|N_{23}|\ll1$).} This phenomenon was first noticed in
Ref.~\cite{KKS}. It is interesting to note that an analogous $\tan\beta$
enhancement also occurs in the $b\to s\gamma$ amplitude \cite{bsgamma},
although its effect is somewhat obscured by possible strong cancellations
against the QCD correction factor.

The results of the calculation in the no-scale and dilaton cases are plotted in
 Figs.~1a,1b respectively, against the gluino mass, for the indicated values
of $m_t$.\footnote{The choice $m_t=170\GeV$ has not been shown because it is
subjected to strict constraints from the $\epsilon_1$ electroweak parameter
\cite{bsg-eps}.} As anticipated, the values of $\tan\beta$ increase as the
corresponding curves move away from the zero axis. Note that $a^{susy}_\mu$
drops off faster than naively expected (\ie, $\propto1/m_{\tilde g}$) since
the $U_{12}$ mixing element decreases as the limit of pure wino and Higgsino
is approached for large $m_{\tilde g}$. Note also that $a^{susy}_\mu$ can have
either sign, in fact, it has the same sign as the Higgs mixing parameter
$\mu$.\footnote{For comparison with earlier work, our sign convention for $\mu$
is opposite to that in Ref.~\cite{HK}.} The incorrect perception that
$a^{susy}_\mu$ is generally negative appears to be based on several model
analyses where either $\mu$ was chosen to be negative or only some of the
neutralino-smuon pieces were kept (which are mostly negative). Interestingly,
the largest allowed values of $\tan\beta$ do not exceed the $a_\mu$ constraint
since consistency of the models (\ie, the radiative breaking constraint)
requires larger gluino masses as $\tan\beta$ gets larger.

Comparing the results shown in Fig. 1 with the allowed ranges in
Eq.~(\ref{bounds}), it is clear that some points in parameter space are already
excluded, depending on the confidence level one wants to use (the dashed lines
in Fig. 1 represent the 95\%CL limit). The corresponding
excluded ranges in the other sparticle masses can be deduced from the
proportionality coefficients given in Table~\ref{Table1}. To show in a
more clear way which region of parameter space is excluded by the present
data, in Fig.~2 we show all the allowed points in parameter space of the two
models (dots and crosses) in the $(m_{\chi^\pm_1},\tan\beta)$ plane for fixed
values of $m_t$. Those points marked with crosses are excluded by the $a_\mu$
constraint at the 95\%CL. Note that for not too small values of $\tan\beta$,
chargino masses in the range $45-120\GeV$ are already excluded; there are no
constraints for $\tan\beta\lsim8$. Using Table~\ref{Table1} this
reach in chargino masses translates into $m_{\tilde q,\tilde g}\approx430\GeV$,
$m_{\tilde e_L,\tilde\mu_L}\approx130\GeV$,
$m_{\tilde e_R,\tilde\mu_R}\approx75\GeV$, $m_{\chi^0_1}\approx60\GeV$, and
$m_{\chi^0_2}\approx120\GeV$.

It is hard to tell what will happen when the E821 experiment reaches its
designed accuracy limit. However, one point should be quite clear, the
supersymmetric contributions to $a_\mu$ can be so much larger than the present
hadronic uncertainty ($\approx\pm1.76\times10^{-9}$) that the latter is
basically irrelevant for purposes of testing a large fraction of the allowed
parameter space of the models. This is not true for the electroweak
contribution and will also not hold for small values of $\tan\beta$. Should the
actual measurement agree very well with the standard model contribution, then
either $\tan\beta\sim1$ or the sparticle spectrum would need to be in the TeV
range. This situation is certainly a window of opportunity for sparticle
detection before LEPII starts operating. Moreover, a significant portion of the
explorable parameter space (those points with $m_{\chi^\pm_1}\gsim100\GeV$
and equivalently $m_{\tilde g}\gsim350\GeV$) is in fact beyond the reach of
LEPII.

Now let us turn briefly to the tau $g-2$ factor. As expected, one generally
obtains $a^{susy}_\tau\sim(m_\tau/m_\mu)^2\,a^{susy}_\mu$ since the dominant
term (the ``gauge-Yukawa" chargino-sneutrino term) is now proportional to
$m^2_\tau$, everything else being the same. The largest values obtained this
way do not exceed $\approx1.6\times10^{-5}$ (and these even occur for points
in parameter space already excluded by the $a^{susy}_\mu$ constraint). In
comparison, the SM contribution has been estimated to be
$a^{SM}_\tau=117.73(0.03)\times10^{-5}$ \cite{SLM}, and the supersymmetric
contribution could easily exceed the present theoretical uncertainty in
$a^{SM}_\tau$. However, the values of $a^{susy}_\tau$ are below the possible
experimental accuracy reachable at hadron supercolliders ($4\times 10^{-5}$
\cite{taug,SLM}) and thus undetectable in the foreseeable future.

\section{Conclusions}
We have computed the supersymmetric contribution to the anomalous magnetic
moment of the muon in the context of $SU(5)\times U(1)$ supergravity models.
The predictions are quite sharp since they depend on only three parameters,
one of which is the top-quark mass. Moreover, the large values of $\tan\beta$,
which are typical in this class of models, enhance the supersymmetric
contribution so much that non-negligible constraints on the parameters of the
models exist even with the present data, and in light of the LEP lower bounds
on the sparticle masses. These contributions are generally much larger than the
electroweak contribution and the present standard model hadronic uncertainty,
and thus should be readily observable at the new E821 Brookhaven experiment.
The potential for decisive exploration of the parameter space of these models
is extremely bright and much greater than the direct experimental production of
sparticles at present and near future collider facilities. We expect that the
qualitative results in this paper will remain valid in a more general class of
supersymmetric models, as long as no new light particles are introduced, and
large values of $\tan\beta$ are allowed. In contrast, in the minimal $SU(5)$
supergravity model one would not expect large contributions to $a^{susy}_\mu$
since the constraint from proton decay requires heavy slepton masses and
$\tan\beta\lsim5$ \cite{minsu5}. Indeed, we find
$|a^{susy}_\mu|\lsim0.2\times10^{-9}$, which is unobservable even for the new
Brookhaven experiment. Finally, experimental limitations indicate that the
supersymmetric contribution to $(g-2)_\tau$ is likely to remain undetected in
the foreseeable future.

\section*{Acknowledgements}
This work has been supported in part by DOE grant DE-FG05-91-ER-40633. The work
of J.L. has been supported by an SSC Fellowship. The work of X.W. has been
supported by a World-Laboratory Fellowship.


\vspace{1cm}
\noindent{\large\bf Figure Captions}
\begin{description}
\item Figure 1: The supersymmetric contribution to the muon anomalous magnetic
moment in (a) the no-scale and (b) the dilaton flipped $SU(5)$ supergravity
models, plotted against the gluino mass for the indicated values of $m_t$
and $\tan\beta$ (which increase in steps of two). The dashed lines represent
the 95\%CL experimentally allowed range.
\item Figure 2: The allowed parameter space of (a) the no-scale and (b) the
dilaton flipped $SU(5)$ supergravity models (in the
$(m_{\chi^\pm_1},\tan\beta)$ plane) for the indicated values of $m_t$. The
points marked by crosses violate the present experimental constraint on
$a_\mu$ at the 95\%CL.
\end{description}
\end{document}